# Oxygen-based digital etching of AlGaN/GaN structures with AlN as etch-stop layers

Jingyi Wu[1,2], Siqi Lei[1,2,3], Wei-Chih Cheng[1,2,4], Robert Sokolovskij[1,2,5], Qing Wang[1], Guangrui (Maggie) Xia [1,6,a)], Hongyu Yu [1,7,8,a)]

[1] School of Microelectronics, Southern University of Science and Technology, Shenzhen, China
[2] Department of Electric and Electronics Engineering, Southern University of Science and Technology, Shenzhen, China
[3] Harbin Institute of Technology, China
[4] Department of Electronic and Computer Engineering, Hong Kong University of Science and Technology, Hong Kong, China
[5] Department of Microelectronics, Delft University of Technology, Delft, Netherlands
[6] Department of Materials Engineering, the University of British Columbia, Vancouver, BC, Canada
[7] GaN Device Engineering Technology Research Center of Guangdong, Southern University of Science and Technology, Shenzhen, China
[8] The Key Laboratory of the Third Generation Semiconductors, Southern University of Science and Technology, Shenzhen, China

[a)] Electronic mail: gxia@mail.ubc.ca, yuhy@sustech.edu.cn

$O_2$-plasma-based digital etching of $Al_{0.25}Ga_{0.75}N$ with a 0.8 nm AlN spacer on GaN was investigated. At 40 W RF bias power and 40 sccm oxygen flow, the etch depth of $Al_{0.25}Ga_{0.75}N$ was 5.7 nm per cycle. The 0.8 nm AlN spacer layer acted as an etch-stop layer in 3 cycles. The surface roughness improved to 0.33 nm after 7 digital etch cycles. Compared to the dry etch only approach, this technique causes less damages. Compared to the selective thermal oxidation with a wet etch approach, this method is less demanding on the epitaxial growth and saves the oxidation process. It was shown to be effective in precisely controlling the AlGaN etch depth required for recessed-AlGaN HEMTs.

# I. INTRODUCTION

Wide bandgap GaN-based high electron mobility transistors (HEMTs) and field-effect transistors (FETs) are able to provide higher breakdown voltage and higher electron mobility than conventional Si-based high power devices[1-3]. Normally-off GaN HEMTs are greatly needed to lower power and to simplify circuit and system architecture, which is one of the major challenges in GaN HEMT technology[2-4]. A recessed-AlGaN/GaN structure is one of the useful options to enable the normally-off operation. Due to the high requirements on the gate dielectric thickness, uniformity, quality and surface morphology, recess etch is a very critical step in the fabrication of GaN HEMT[2, 4].

Reactive ion etching (RIE) with chlorine ($Cl_2$) or boron trichloride ($BCl_3$) plasma gives high selectivity between AlGaN and GaN. However, they cause damages to AlGaN[3-6], which can significantly affect the uniformity, surface quality and morphology. Therefore, $Cl_2$ and $BCl_3$ etching recipes are not suitable for the very shallow recess etch[3, 6]. Another well-known method contains a selective thermal oxidation step of AlGaN over GaN at 500 to 600 °C and a wet etch in a potassium hydroxide (KOH) solution[7-9]. In another work, a GaN cap layer was used as the oxidation mask and another GaN as the etch-stop layer[7-9]. This method can generate very smooth surfaces with root mean square RMS smoothness below 0.3 nm. However, it is limited for practical application due to the more complicated epitaxy and extra thermal oxidation steps.

Oxygen plasma based digital etching is the third etch option. A cycle of digital etching includes a plasma oxidation step and a wet etching step to oxidize the target material and to remove the oxidation products. Compared to the $Cl_2$ or $BCl_3$ plasma etch, it produces better surface quality. Also, it does not require very complicated epitaxy as the second recess etch method discussed above. Theoretically, the oxygen penetration depth depends on the radio frequency (RF) bias power of the plasma. Therefore, the thickness of oxidation and the amount etched in one cycle saturate with the oxidation time at a certain temperature and plasma power [10-11]. This phenomenon realizes the self-limiting recess etching process.

A digital etching recipe containing oxygen ($O_2$) plasma was first shown by S. D. Burnham et al. in 2010 [10]. Normally, nitrous oxide ($N_2O$) or $O_2$ are used as the effective gases in the oxidation step. A dry $BCl_3$ plasma etching and/or a wet etch in a hydrochloric acid (HCl) solution are used to etch away the oxides formed in the oxidation step[3-4, 10-13]. Digital etching is commonly used in etching group III-V semiconductors such as gallium arsenide (GaAs) [6, 11-16]. P-GaN HEMTs made using digital etching was demonstrated recently [16], where 1 nm AlN was used as the etch-stop layer and 60 nm p-GaN was etched. In their work, $N_2O$ plasma was the oxidant, and the RF plasma power was 30 W. However, digital etching of AlGaN has only been studied in limited literature. Previous studies used cyclical $O_2$ and $Cl_2/BCl_3$ plasma[16-17]. One work by R. Sokolovskij et al. used $O_2$ plasma oxidation and HCl wet etching for AlGaN/GaN sensors without any etch-stop layers[17]. They investigated the digital etching behavior with different power settings and equipment. The AlGaN layers studied in the literature were partially recessed, and the recess depth was controlled by the etching cycle numbers [17-20].

This work was designed to study the digital etching of AlGaN with an AlN spacer/etch-stop layer, which is helpful for better controlling the AlGaN recess depth. Meanwhile, improved surface roughness was observed using this method.

## II. EXPERIMENTAL

In this work, oxygen plasma in an inductively coupled plasma (ICP) tool was used as the oxidant. The AlGaN/GaN/Si substrates used in our work were grown and provided by Enkris Semiconductor Inc.. From top to bottom, the epitaxial layers on Si are a 2.5 nm GaN cap layer, a 19 nm $Al_{0.25}Ga_{0.75}N$ barrier layer, a 0.8 nm AlN spacer layer and a 700 nm intrinsic GaN, and buffer layers.

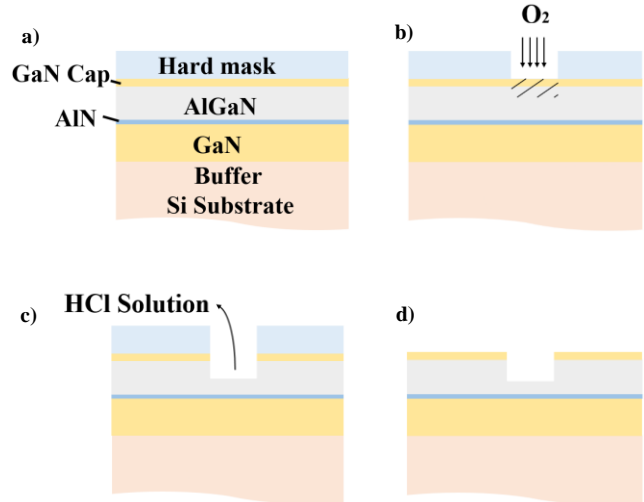

FIG. 1. Schematics showing the cross-sections of the epitaxy structure (a) before etching, (b) during the ICP oxidation (dry etching), (c) after the HCl wet etching, d) after the hard mask was etched away by a BOE solution for etch depth measurements.

As shown in Fig. 1. (a), a layer of 100 nm silicon dioxide ($SiO_2$) acted as the hard mask in the dry etch, which was grown by plasma enhanced chemical vapor deposition (PECVD) using $N_2O$. Its growth rate was 0.9 nm per second on average at 350 °C. The pattern of the hard mask was defined using photolithography and fluorine based ICP etching to expose the GaN cap surface and the underlying AlGaN.

The wafer pieces were divided into five groups, each of which saw three to seven cycles of digital etching respectively. The flow diagrams of the digital etching are shown in Fig. 1.

The first step of the digital etching was to oxidize AlGaN by $O_2$ plasma with an ICP power of 450 W and RF bias power of 40 W. The $O_2$ flow rate was 40 sccm, and no other gases were used. The chamber pressure used was at 8 mTorr. We did a series of test under 75 and 40W (Fig. 2.). In the wet etch steps, a 90 sec etch in an HCl solution (deionized water: HCl = 5:1) was used. The AlGaN etch depth was measured by atomic force microscopy (AFM). On each sample, the depth at six points was measured and the average etch depth was used. The etch depth measurement error was 1 nm. The hard mask and the native oxide of the GaN cap layer were etched away by BOE solutions before AFM measurements.

For each cycle, it was essential to obtain the etch depth dependence on the oxidation time. A series of samples were oxidized with 2 to 5 min oxidation time and under two RF power settings of 40 and 75 W. In Fig. 2. (a), the oxidation depth saturates at 3 and 4 min with 40 W and 75 W RF power respectively, showing the self-limiting behavior. According to the theories of digital etch[2, 17], ideally, oxygen would not penetrate deeper beyond a critical time point, which was selected as our oxidation time for good repeatability. The oxidation depth of 9 nm/cycle at 75 W was too fast to be controlled for 19 nm $Al_{0.25}Ga_{0.75}N$ etching. Therefore, the recipe of 3 min oxidation time with 40 W RF power was used in our experiments.

## III. RESULTS AND DISCUSSION

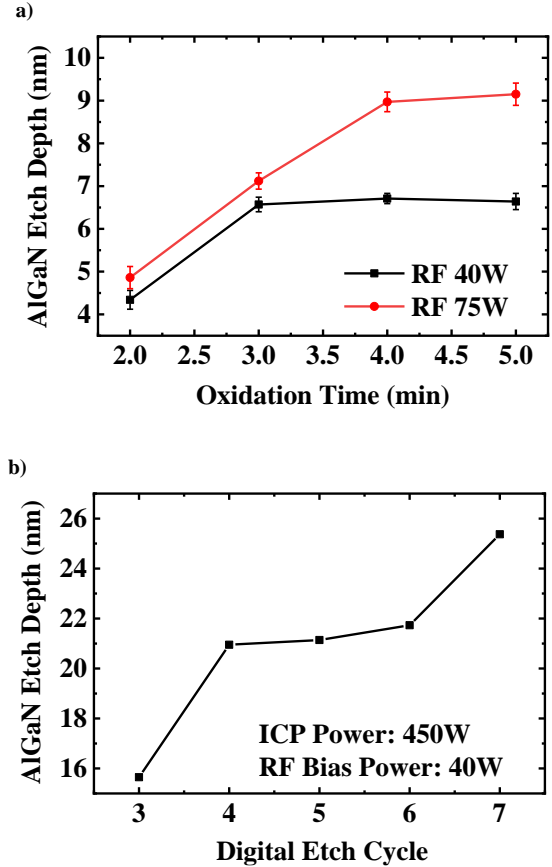

FIG. 2. (a) Oxidation depth under different RF bias power settings in one cycle. It can be seen that the oxidation depth saturates after 3 and 4 min with 40 W and 75 W RF power respectively. The etch depth was 6.3 and 9.4 nm per cycle respectively for these two power settings. (b) Etch depth vs. the cycle number using 3 min oxidation time in each cycle at 40 W RF power.

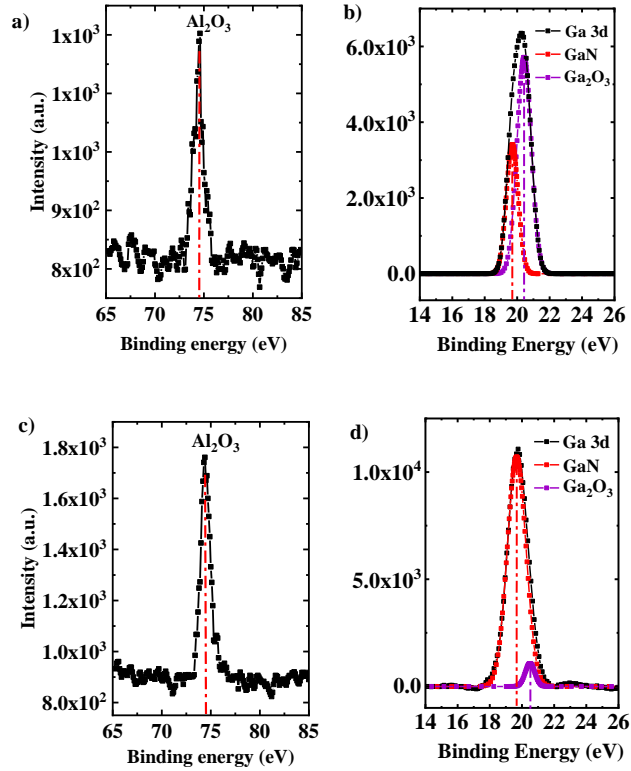

FIG. 3. The X-ray photoelectron spectroscopy measurement of samples after (a), (b) the 4th cycle's and (c), (d) the 5th cycle's oxidation in ICP. It should be noted that the oxides in the 4th cycle were products from the $Al_{0.25}G_{0.57}N$ oxidation, and in the 5th cycle, those were mainly from the AlN oxidation.

As seen in Fig. 2. (b), the etching depth reached a plateau at the 4th cycle, and then continued after the 6th cycle. Oxygen plasma reacted with $Al_{0.25}Ga_{0.75}N$ and AlN, and the oxidation products were investigated by X-ray photoelectron spectroscopy (XPS). After the 4th and the 5th plasma oxidation step, before dipping in the HCl solution, the XPS measurement results proved that the oxidation products had aluminum oxide $Al_2O_3$ and gallium oxide $Ga_2O_3$ (Fig. 3) [21-22]. It was clear that the proportion of $Ga_2O_3$ decreased while the peak of $Al_2O_3$ was much more distinct after the 5th oxidation step. The four chemical reactions of the oxidation and wet etching are described as below:

$$2AlGaN+3O_2 \rightarrow Ga_2O_3+Al_2O_3+N_2$$

$$4AlN+3O_2 \rightarrow 2Al_2O_3+2N_2$$

$$Ga_2O_3+6HCl \rightarrow 2GaCl_3+3H_2O$$

$$Al_2O_3+6HCl \rightarrow 2AlCl_3+3H_2O$$

Digital etching paused after 4 etching cycles as shown in Fig. 2. (b) when the etching of AlGaN was completed. After the 4th cycle, the etch depth was 20.53 nm measured by AFM. It means that all the $Al_{0.25}Ga_{0.75}N$ layer was etched. In the 5th cycle, the AlN barrier layer was oxidized and became a layer of $Al_2O_3$ on GaN. However, the diluted HCl solution had too low an etching rate to remove $Al_2O_3$ in 1.5 min[11, 14]. Therefore, the etching seems to be paused by this $Al_2O_3$ layer. After the 6th cycle, the etching continued as $Al_2O_3$ was etched by more HCl solution etching time and the oxygen plasma bombardment. After the 6th cycle, the thick GaN layer was exposed to oxygen in ICP, and was oxidized to $Ga_2O_3$, which could be dissolved in the HCl solution. Therefore, the etching depth continued to increase after the 6th cycle.

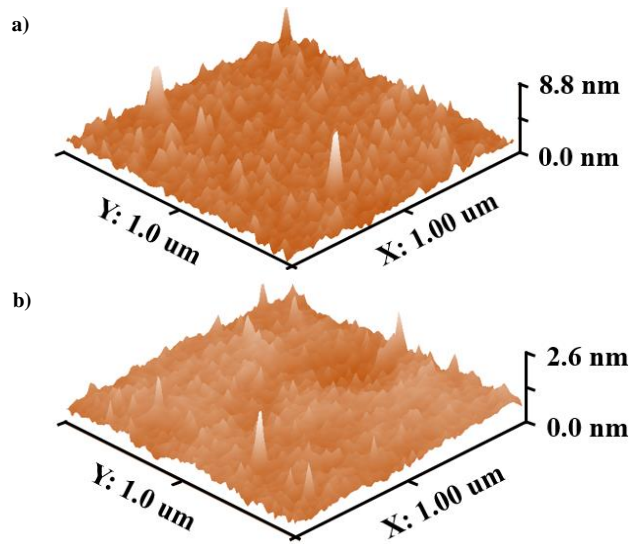

FIG. 4. AFM images showing 1 μm × 1 μm area after (a) 3 cycles, and (b) 7 cycles of digital etching.

TABLE I. Average roughness values measured on six 1μm × 1μm areas before etching and after etch cycle 3 to 7.

| Cycle # | RMS roughness | Mean roughness |
| --- | --- | --- |
| No etch | 0.855 nm | 0.855 nm |
| 3 | 0.656 nm | 0.465 nm |
| 4 | 0.618 nm | 0.446 nm |
| 5 | 0.472 nm | 0.351 nm |
| 6 | 0.563 nm | 0.377 nm |
| 7 | 0.330 nm | 0.263 nm |

The surface morphology of the AlGaN/GaN samples after 3 and 7 cycles of digital etching process was compared in Fig. 4, which were measured by AFM on six areas per sample with a scan area of 1 × 1 μm$^2$. Each surface was characterized using a root-mean-square (RMS) roughness and a mean roughness (Table I). The roughness decreased slightly with increasing number of cycles. However, the RMS roughness of the 6$^{th}$ cycle increased slightly. Due to the very thin thickness of AlN (0.8 nm), the epitaxial growth of this layer resulted in a higher AlN thickness non-uniformity and hence higher $Al_2O_3$ thickness non-uniformity. After the 7$^{th}$ cycle, the RMS improved from 0.656 to 0.330 nm.

## IV. SUMMARY AND CONCLUSIONS

In summary, O$_2$-plasma-based digital etching of Al$_{0.25}$Ga$_{0.75}$N with 0.8 nm AlN spacer on GaN was investigated using an ICP etcher. At 40 W RF bias power and 40 sccm oxygen flow, the etch depth of Al$_{0.25}$Ga$_{0.75}$N was 5.7 nm per cycle. The 0.8 nm AlN spacer layer acted as an etch-stop layer for AlGaN recess. The surface roughness improved after the digital etch cycles to 0.330 nm. Compared to the dry etch only approach, this technique causes less damages. Compared to the selective oxidation with a wet etch approach, this method is less demanding on the epitaxial growth and saves the oxidation process. The presence of AlN as an etch-stop layer for the digital etch guarantees the better recess control for the preparation of gate-recessed HEMT.

## ACKNOWLEDGMENTS

This work was supported under "Research of key techniques and applications of GaN-on-Si power device (Grant No: 2019B010128001)" by Guangdong Science and Technology Department, and under "Research of low cost fabrication of GaN power devices and system integration" (Grant No: JCYJ20160226192639004)" funded by Shenzhen Municipal Council of Science and Innovation. The work was conducted at Materials Characterization and Preparation Center (MCPC) at SUSTech, and we acknowledge the technical support from the staff and engineers at MCPC.